\documentclass[aps,prx,preprint,superscriptaddress]{revtex4-1}

\usepackage{graphicx}
\usepackage{amsmath}
\usepackage{color}
\usepackage[dvipsnames]{xcolor}
\usepackage[normalem]{ulem} % to strike out plain text

\newcommand{\CGC}[6]{ \mbox{$ \left( #1 #2 , #3 #4\,|\, #5 #6 \right) $} }
\newcommand{\ME}[3]{ \mbox{$\langle #1\,|\,#2\;|\;#3\rangle $}}
\newcommand{\Skalar}[2]{ \mbox{$\langle\,#1\,|\,#2\,\rangle $} }

\begin{document}

\title{
A new method for measuring angle-resolved phases in photoemission
}

\author{Daehyun You}
\author{Kiyoshi Ueda}
\email[corresponding author;\ ]{kiyoshi.ueda@tohoku.ac.jp}
\affiliation{Institute of Multidisciplinary Research for Advanced Materials, Tohoku University, Sendai 980-8577, Japan}
\author{Elena V. Gryzlova}
\author{Alexei N. Grum-Grzhimailo}
\affiliation{Skobeltsyn Institute of Nuclear Physics, Lomonosov Moscow State University, Moscow 119991, Russia}
\author{Maria M. Popova}
\affiliation{Skobeltsyn Institute of Nuclear Physics, Lomonosov Moscow State University, Moscow 119991, Russia}
\affiliation{Faculty of Physics, Lomonosov Moscow State University, Moscow 119991, Russia}
\author{Ekaterina I. Staroselskaya}
\affiliation{Faculty of Physics, Lomonosov Moscow State University, Moscow 119991, Russia}

\author{Oyunbileg Tugs}
\author{Yuki Orimo}
\affiliation{Department of Nuclear Engineering and Management, Graduate School of Engineering, The University of Tokyo, 7-3-1 Hongo, Bunkyo-ku, Tokyo 113-8656, Japan}
\author{Takeshi Sato}
\author{Kenichi L. Ishikawa}
\affiliation{Department of Nuclear Engineering and Management, Graduate School of Engineering, The University of Tokyo, 7-3-1 Hongo, Bunkyo-ku, Tokyo 113-8656, Japan}
\affiliation{Photon Science Center, Graduate School of Engineering, The University of Tokyo, 7-3-1 Hongo, Bunkyo-ku, Tokyo 113-8656, Japan}
\affiliation{Research Institute for Photon Science and Laser Technology, The University of Tokyo, 7-3-1 Hongo, Bunkyo-ku, Tokyo 113-0033, Japan}
\author{Paolo Antonio Carpeggiani}
\affiliation{Institut f\"ur Photonik, Technische Universit\"at Wien, 1040 Vienna, Austria}
\author{Tam\'as Csizmadia}
\author{Mikl\'os F\"ule}
\affiliation{ELI-ALPS, ELI-HU Non-Profit Ltd., Dugonics t\'er 13, H-6720 Szeged, Hungary}
\author{Giuseppe Sansone}
\author{Praveen Kumar Maroju}
\affiliation{Physikalisches Institut, Albert-Ludwigs-Universit\"at Freiburg, 79106 Freiburg, Germany}
\author{Alessandro D'Elia}
\affiliation{University of Trieste, Department of Physics, 34127 Trieste, Italy}
\affiliation{IOM-CNR, Laboratorio Nazionale TASC, 34149 Basovizza, Trieste, Italy}
\author{Tommaso Mazza}
\author{Michael Meyer}
\affiliation{European X-Ray Free Electron Laser Facility GmbH, Holzkoppel 4, 22869 Schenefeld, Germany}
\author{Carlo Callegari}
\author{Michele Di Fraia}
\author{Oksana Plekan}
\author{Robert Richter}
\affiliation{Elettra -- Sincrotrone Trieste S.C.p.A., 34149 Basovizza, Trieste, Italy}
\author{Luca Giannessi}
\affiliation{Elettra -- Sincrotrone Trieste S.C.p.A., 34149 Basovizza, Trieste, Italy}
\affiliation{INFN - Laboratori Nazionali di Frascati, 00044 Frascati, Rome, Italy}
\author{Enrico Allaria}
\affiliation{Elettra -- Sincrotrone Trieste S.C.p.A., 34149 Basovizza, Trieste, Italy}
\author{Giovanni De Ninno}
\affiliation{Elettra -- Sincrotrone Trieste S.C.p.A., 34149 Basovizza, Trieste, Italy}
\affiliation{Laboratory of Quantum Optics, University of Nova Gorica, Nova Gorica 5001, Slovenia}
\author{Mauro Trov\`o}
\author{Laura Badano}
\author{Bruno Diviacco}
\author{Giulio Gaio}
\affiliation{Elettra -- Sincrotrone Trieste S.C.p.A., 34149 Basovizza, Trieste, Italy}
\author{David Gauthier}
\altaffiliation{Now at LIDYL, CEA, CNRS, Universit\'e Paris-Saclay, CEA-Saclay, 91191 Gif-sur-Yvette, France.}
\affiliation{Elettra -- Sincrotrone Trieste S.C.p.A., 34149 Basovizza, Trieste, Italy}
\author{Najmeh Mirian}
\author{Giuseppe Penco}
\author{Primo\v{z} Rebernik Ribi\v{c}}
\author{Simone Spampinati}
\author{Carlo Spezzani}
\affiliation{Elettra -- Sincrotrone Trieste S.C.p.A., 34149 Basovizza, Trieste, Italy}
\author{Kevin C. Prince}
\email[corresponding author;\ ]{prince@elettra.eu}
\affiliation{Elettra -- Sincrotrone Trieste S.C.p.A., 34149 Basovizza, Trieste, Italy}
\affiliation{Centre for Translational Atomaterials, Swinburne University of Technology, 3122 
Melbourne, Australia}

\date{\today}

\begin{abstract}
Quantum mechanically, photoionization can be fully described by the complex photoionization amplitudes that describe the transition between the ground state and the continuum state. Knowledge of the value of the phase of these amplitudes
has been a central interest in photoionization studies and newly developing attosecond science, since the phase can reveal important information about phenomena such as electron correlation. 
We present a new attosecond-precision interferometric method of angle-resolved measurement for the phase of the photoionization amplitudes, using two phase-locked Extreme Ultraviolet pulses of frequency $\omega$ and $2\omega$, from a Free-Electron Laser. 
Phase differences  $\Delta \tilde \eta$ between one- and two-photon ionization channels, averaged over multiple wave packets, are extracted for neon $2p$ electrons as a function of emission angle 
at photoelectron energies 7.9, 10.2, and 16.6 eV.
$\Delta \tilde \eta$ is nearly constant for emission parallel to the electric vector but increases at 10.2 eV for emission perpendicular to the electric vector.  
We model our observations with both perturbation and \textit{ab initio} theory, and find excellent agreement. 
In the existing method for attosecond measurement, 
Reconstruction of Attosecond Beating By Interference of Two-photon Transitions (RABBITT), 
a phase difference between two-photon pathways involving absorption
and emission of an infrared photon is extracted.
Our method can be used for extraction of a phase difference between 
single-photon and two-photon pathways 
 and provides a new tool for attosecond science, which is complementary to RABBITT.
\end{abstract}

\maketitle

\section{\label{sec:introduction}Introduction}
The  age  of  attosecond  physics  was  ushered  in  by  the  invention  of  methods  for  probing phenomena on a time scale less than femtoseconds~\cite{Krausz2009}. 
A phenomenon occurring on this time scale is photoemission delay. 
When the photon energy is far from resonance,
the photoemission delay for single photon ionization can be associated with the Wigner delay experienced by an electron scattering off the ionic potential~\cite{Wigner1955}. 
Quantum mechanically, the photoionization process is fully described by the complex photoionization amplitudes describing transitions between the ground state and the continuum state. The photoemission delay can be expressed as the energy derivative of the phase of the photoionization amplitude, and therefore measuring the photoemission delay and the energy-dependent phase of the photoionization amplitude are practically equivalent. Their measurement
is one of the central interests in attosecond science~\cite{Pabst2016,Klunder2011,Guenot2014,Guenot2012,Palatchi2014,Kheifets2013,Saha2014,Argenti2017,Busto2019,Vos2018,Azoury2019,Fuchs2020}, because they are a fundamental probe of the photoionization process and can reveal important information about, for example, electron-electron correlations (see, e.g.~\cite{Ossiander2017}). 

Currently two methods are available to measure these quantities: streaking and RABBITT (Reconstruction of Attosecond Beating By Interference of Two-photon Transitions), both of which require the use of an IR dressing field. 
We present a new interferometric method of angle-resolved measurement for the photoionization phase, using two phase-locked Extreme Ultraviolet (XUV) pulses of frequency $\omega$ and $2\omega$, from a Free-Electron Laser (FEL), without a dressing field.

In attosecond streaking~\cite{Schultze2010}, an ultrafast, short-wavelength pulse ionizes an electron, and a
femtosecond infrared (IR) pulse acts as a streaking field, by changing
the linear momentum of the photoelectron. 
In this technique, one can extract the photoemission delay difference between two photoemission lines at two different energies, arising for example from two different subshells~\cite{Schultze2010} or the main line and satellites~\cite{Ossiander2017}. 
Generally, time-of-flight electron spectrometers located in the streaking direction (the direction of linear polarization) are used, so that this method does not give access to angular information. A related method is the attosecond clock technique~\cite{Eckle2008sci,Eckle2008nat,Pfeiffer2012,Pfeiffer2011},
in which 
streaking by the circularly polarized laser pulse is in the angular direction.

The second technique for measuring 
photoemission delays, RABBITT, is interferometric: it uses a
train of attosecond pulses dressed by a phase-locked IR pulse~\cite{Paul2001}.
In the
RABBITT technique, the phase difference between a pair of two-photon pathways whose final energy is separated by multiples of an infrared photon energy is extracted. The extracted value is related to the phase
difference of the two-photon ionization amplitudes at the pair of energies. 
For two energy points separated by twice the IR photon energy, the phase difference divided by twice the IR photon energy can be regarded as
a finite difference approximation to the energy derivative of phase of the two-photon ionization amplitude.
The pulse duration requirements are relaxed: for
example, pulse trains and IR pulses of 30 fs duration may be used~\cite{Guenot2012}. Usually the IR pulse is the
fundamental of the odd harmonics in the pulse train,
although \textcite{Loriot2017} reported a variant
using the second
harmonic. Recent work on phase retrieval includes methods
based on photo-recombination~\cite{Azoury2019, Azoury2019a}, two-color, two-photon ionization via a resonance~\cite{Villeneuve2017}, and a proposal to use successive harmonics of circularly polarized light~\cite{Donsa2019}.

The phase of the photoionization amplitude depends on photoelectron energy $\epsilon$  
and it may also depend on the electron’s emission direction.
There is a physical origin for the directional anisotropy of the
amplitude: an electron wave packet may consist of two or more partial waves, with different angular momenta and phases.
There has been significant theoretical work on the angle-dependent time delay, for example~Ref. [\onlinecite{Dahlstrom2014,Watzel2015, Heuser2016,Mandal2017,Ivanov_Kheifets2017,Bray2018,Banerjee2019}], but fewer related experimental reports~\cite{Heuser2016,Cirelli2018,Vos2018}, all using the RABBITT technique. 
The Wigner delay is theoretically isotropic for single-photon ionization of He, but \textcite{Heuser2016} observed an angular dependence in photoemission delay, attributed to the XUV+IR two-photon ionization process, inherent in RABBITT interferometry.

In the present work, we demonstrate interferometric measurements of the relative phase of 
single-photon and two-photon ionization amplitudes.
The interference is created between a two-photon ionization process driven by a fundamental wavelength, and a single-photon ionization process driven by its phase-locked, weaker, second harmonic, in a setup like that demonstrated at visible wavelengths~\cite{Shapiro2003}. Using short-wavelength, phase-locked XUV light, we measure angular
distributions of photoelectrons emitted from neon, and determine the 
phase difference for one- and
two-photon ionization wavepackets.
The extremely short (attoseconds) pulses required for streaking or attosecond pulse trains for RABBITT are not needed, and instead access to 
photoemission phase with attosecond precision is provided by optical phase control with precision of a few attoseconds, which is available from the Free-Electron Laser FERMI~\cite{Prince2016}.

The rest of the manuscript is structured as follows: in Section~\ref{sec:notation} we introduce the necessary notation and the basic processes that may be active in the experiment; in Section~\ref{sec:experimental} and~\ref{sec:theory} we describe respectively the experimental and theoretical methods used. In Section~\ref{sec:Results} we present and compare experimental and theoretical results. We discuss  in Section~\ref{sec:Discusson} the relationship between our data, namely the angular distribution of photoelectrons created by collinearly polarized biharmonics, and the time-delay studies described in the introductory section. Section~\ref{sec:summary} presents our summary and outlook, 
and the Appendix gives details of the derivation of some equations.

\section{\label{sec:notation}Notation and basic processes}
We use Hartree atomic units unless otherwise stated, and spherical coordinates
$\textbf{r}= \{r,\theta,\varphi\}$ relative to the direction of polarization of the bichromatic field (linear horizontal in the experiment). We assume the electric dipole approximation, and the experiment is cylindrically symmetric about the electric vector, so that there is no dependence on the azimuthal angle $\varphi$. The bichromatic electric field is described by:
\begin{equation}
\label{eq:Efield}
    E\left( t \right)
        = \sqrt{I_\omega \left( t \right)} \cos \omega t
        + \sqrt{I_{2\omega} \left( t \right)} \cos \left( 2\omega t  - \phi \right) \,,
\end{equation}
where $\omega$ and $2\omega$ are angular frequencies, $I_\omega (t)$ and $I_{2\omega} (t)$ are the pulse envelopes,
$\phi$ denotes the $\omega$-$2\omega$ relative phase.

We can consider the experimental sample as an ensemble of identical atoms of infinitesimal size, so we can reduce the theoretical treatment to that of a single atom centered at the coordinates' origin. The general form (omitting as implicit the dependence on $\theta,\varphi$) of an electron wave packet sufficiently far away from the origin is:
%
%\begin{equation}
%\label{eq:wavepacket}
%	\int_0^\infty \,c \left( \epsilon \right) e^{-i \epsilon t} e^{i \left[ \sqrt{2\epsilon} r + \eta \left( \epsilon \right) \right] } d\epsilon,
%\end{equation}
%

\begin{equation}
\label{eq:wavepacket}
\int_0^\infty c(\epsilon) \, e^{-i \epsilon t} e^{i\left[\sqrt{2 \epsilon} \, r + 
f(r,\epsilon) + \eta(\epsilon) \right]} d \epsilon \,,
\end{equation}
where $\epsilon$ is the photoelectron kinetic energy, $c(\epsilon)$ the real-valued
amplitude, $\eta(\epsilon)$ the phase, and the term
$f(r, \epsilon) = \frac{Z}{\sqrt{2 \epsilon}} \ln \sqrt{8 \epsilon} \, r$ accounts
for the Coulomb field of the residual ion with charge $Z$. In our case $Z=+1$.

In the $\omega$-$2\omega$ process, i.e., one driven by the field in Eq.~(\ref{eq:Efield}), the wave packet can be expressed as
%
%\begin{equation}
%\label{eq:wavepacket2}
%	\int_0^\infty \, e^{-i \epsilon t} e^{i \sqrt{2 \epsilon} r} \left\{ c_\omega(\epsilon) e^{i \eta_\omega(\epsilon)} + c_{2\omega}(\epsilon) e^{i \left[ \eta_{2\omega} (\epsilon) + \phi \right] } \right\} d\epsilon\;.
%\end{equation}

\begin{equation}
\label{eq:wavepacket2}
\int_0^\infty c(\epsilon) \, e^{-i \epsilon t} 
e^{i \left[\sqrt{2 \epsilon} \, r + 
	f(r,\epsilon) \right]}
\{ c_{\omega}(\epsilon) e^{i \eta_{\omega}(\epsilon)}
+  c_{2\omega}(\epsilon) e^{i [ \eta_{2\omega}(\epsilon) + \phi]}
      \} d \epsilon \,.
\end{equation}

The photoelectron yield as a function of optical phase $\phi$ (we omit the spatial coordinates on the right-hand side) is given by
\begin{eqnarray} \label{eq:ii}
    I(\theta,\varphi;\phi) \doteq I(\theta;\phi) & = & \int_0^\infty c_\omega(\epsilon)^2 + c_{2\omega}(\epsilon)^2 + 2c_\omega(\epsilon) c_{2\omega}(\epsilon) \cos \left[ \phi - \Delta \eta \left( \epsilon \right) \right] \, d\epsilon \nonumber \\
    \label{eq:yield}
     & &\equiv A_0 + A \cos \left[ \phi - \Delta \eta \left( \bar{\epsilon} \right) \right],
\end{eqnarray}
where
$\bar{\epsilon}$ is the average kinetic energy of the wave packet, and
$\Delta \eta \left( \epsilon \right) \equiv \eta_{\omega}(\epsilon) - \eta_{2\omega}(\epsilon)$
is the phase of the two-photon ionization relative to the single-photon ionization.

This treatment may be generalized to the case of multiple wave packets, that is to say, with more than one magnetic quantum number $m$ of the residual ion. 
Wave packets with each value of $m$ interfere separately, and then incoherently add.
In particular,
expressing the photoionization yield as in Eq.~(\ref{eq:yield})
\begin{equation} \label{eq:mod}
	I(\theta;\phi) = A_0 + A \cos \left( \phi -  \Delta \tilde \eta \right) = \sum_m [A_{0,m} + A_m \cos (\phi - \Delta \eta_m)],
\end{equation}
where summation is over the wave packets, leading to

\begin{equation} \label{eq:summation}
	A_0 = \sum_m A_{0,m}, \qquad Ae^{i\Delta \tilde \eta} = \sum_m A_m e^{i\Delta \eta_m}.
\end{equation}

The second equation defines an average phase difference $\Delta \tilde \eta$ of $\{\Delta\eta_m\}$, weighted in terms of the corresponding phase factors. 
Eqs.~(\ref{eq:ii}) and (\ref{eq:mod})
indicate that the yield of photoelectrons emitted by a bichromatic pulse in a particular direction oscillates sinusoidally as a function of the optical phase $\phi$.

\section{\label{sec:experimental}Experimental Methods and Setup}
The experimental methods have been described elsewhere \cite{Prince2016} and here we summarise the main aspects, and the parameters used.
The experiment was carried out at the Low Density Matter Beamline~\cite{Lyamaev2013,Svetina2015} of the FERMI Free-Electron Laser~\cite{Allaria2015}, using the Velocity Map Imaging (VMI) spectrometer installed there. The VMI measures the projection of the Photoelectron Angular Distribution (PAD) onto the planar detector (horizontal); the PAD is obtained as an inverse Abel transform of this projection, using the BASEX method~\cite{BASEX}.  
The images were divided into two halves along the line of the electric vector, labelled ``left" and ``right", and analysed separately. The PADs from the two halves agreed generally, but the detector for the right half showed a small non-uniformity in detection efficiency. Therefore the PADs were analysed using the left half of the detector, denoted as 0 - 180$^{\circ}$ below.

The sample consisted of a mixture of helium and neon, and the helium PAD was used to calibrate the phase difference between the $\omega$ and $2\omega$ fields. The atomic beam was produced by a supersonic expansion and defined by a conical skimmer and vertical slits. The length of the interaction volume along the light propagation direction was approximately 1~mm. In other experiments~\cite{Guenot2014,Palatchi2014}, use of two gases allowed referencing of
the photoemission delay of one electron to that of another.
In the present case, we used the admixture of helium to provide a phase reference. When the Free-Electron Laser wavelength is changed, the mechanical settings of the magnetic structures (undulators) creating the light are changed. This may introduce an unknown phase error between fundamental and second harmonic light. We have recently shown that the PAD of helium $1s$
electrons can be used to determine the absolute optical phase difference between the $\omega$ and $2\omega$ fields, with input of only few theoretical parameters~\cite{DiFraia2019}. 

\begin{figure}[ht]
    \centering
\includegraphics[trim = {0 220 0 0},clip, width=5in]{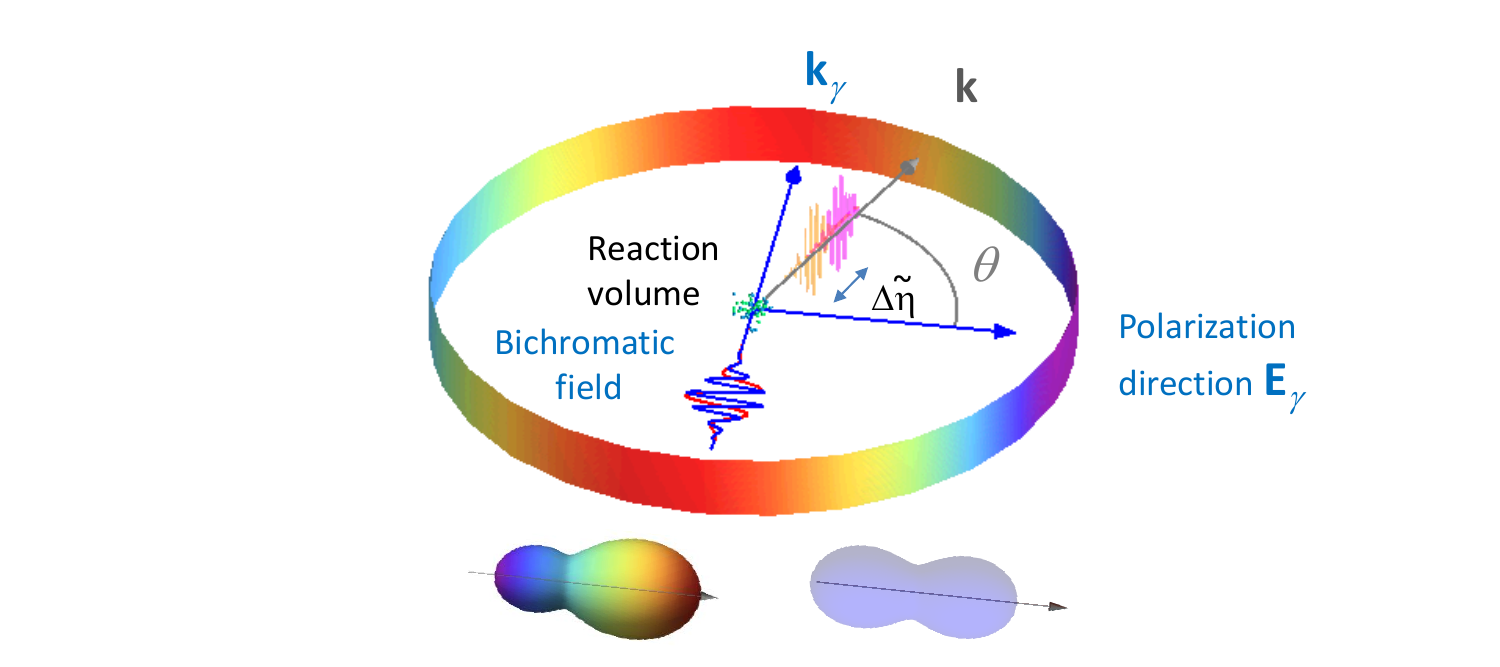}
    \caption{\label{fig:1}Scheme of the experiment: bichromatic, linearly polarized light (red and blue waves),
with momentum ${\bf k}_\gamma$ and electric vector ${\bf E}_\gamma$,
ionizes neon in the reaction volume. The electron wave packets (yellow and magenta waves) are emitted with electron momentum ${\bf k}$.
The $m$-averaged phase difference $\Delta \tilde \eta$ between wavepackets created by one- and two-photon ionization depends on emission angle.
The photoelectron angular distribution depends on the relative (optical) $\omega$-$2\omega$  phase $\phi$.
Lower figures: polar plots of photoelectron intensity 
at $E_{\rm k} = 16.6$~eV for coherent harmonics (asymmetric, coloured plot) and incoherent harmonics (symmetric, grey plot).
}
\end{figure}

The light beam consisted of two temporally overlapping harmonics with controlled relative phase $\phi$, Eq.~(\ref{eq:Efield}), and
irradiated the sample, as shown schematically in {Fig.~\ref{fig:1}}.
The  intense fundamental radiation caused two-photon
ionization, while the weak second harmonic gave rise to single-photon
ionization. The energies of the photoelectrons created coherently in the
two channels are identical, and electrons with the same linear
momentum interfere~\cite{Villeneuve2017}. The PAD $I\left(\theta; \phi \right)$ was measured as a function of the phase $\phi$; from the component oscillating with $\phi$, the scattering phases were extracted, as shown in Section~\ref{sec:Results}. The wavelength was then changed and the measurement repeated. 

The relative phase of the two wavelengths was controlled by means of the electron delay line or phase shifter~\cite{Diviacco2011,Prince2016} used previously. It has been calculated that the two pulses have good temporal overlap with slightly different durations and only a small mean variation of the relative phase of two wavelengths within the Full Width at Half Maximum of the pulses, for example 0.07~rad for a fundamental photon energy of 18.5~eV~\cite{Prince2016}.

The intensities of the two wavelengths for the experiments were set as follows. With the last undulator open (that is, inactive), the first five undulators were set to the chosen wavelength of the first harmonic. A small amount of spurious second harmonic radiation (intensity of the order 1\% of the fundamental) is produced by the undulators~\cite{Giannessi2018}, and to absorb this, the gas filter available at FERMI was filled with helium. 
Helium is transparent at all of the fundamental wavelengths used in this study. The two-photon photoelectron signal from the neon and helium gas sample was observed with the VMI spectrometer. The last undulator was then closed to produce the second harmonic and the photoelectron spectrum of the combined beams was observed. The single-photon ionization by the second harmonic is at least an order of magnitude stronger than the two-photon ionization by the fundamental. The helium gas pressure in the gas filter was then adjusted to achieve a
ratio of the ionization rates due to two-photon and single-photon ionization of 1:2 for kinetic energies of 7.0 and 10.2 eV. For the kinetic energy of 15.9 eV, the ratio was set to 1:4. The bichromatic beam was focused by adjusting the curvature of the Kirkpatrick-Baez active optics~\cite{Zangrando2015}, and verified experimentally by measuring the focal spot size of the second harmonic with a Hartmann wavefront sensor. This instrument was not able to measure the spot size of the beams at the fundamental wavelengths, so it was calculated~\cite{Raimondi2013}. The measured spot was elliptical with a size $(4.5 \pm 1) \times (6.5 \pm 1)$ $\mu$m$^2$ (FWHM), and the estimated pulse duration was 100~fs. 

Table \ref{tab:Table1} summarizes the experimental parameters: fundamental photon energy ($\hbar \omega$), kinetic energy ($E_\mathrm{k} = 2 \hbar \omega - 21.6$~eV) of the Ne photoelectrons emitted via single-photon (2$\omega)$ or two-photon ($\omega+\omega$) ionization, average pulse energy of the first harmonic at the source and at the sample, beamline transmission, and average irradiance at the sample calculated from the above spot sizes and pulse durations.
\begin{table}[h]
\caption{\label{tab:Table1} Experimental parameters. $E_\mathrm{k}$ is the kinetic energy of the Ne $2p$ photoelectrons. The pulse energies at the sample were calculated by multiplying the values at the source by the transmission of the beamline~\cite{Svetina2015}, which takes account of reflection and geometric losses. Pulse energies and irradiance are those of the fundamental.
}
    \centering
    \begin{tabular*}{\textwidth}{@{}l*{15}{@{\extracolsep{0pt plus12pt}}l}}
        \hline
        $\hbar \omega$, eV & $E_\mathrm{k}$, eV & Pulse energy, $\mu$J & Beamline & Pulse energy, $\mu$J & Average irradiance \\
        & & (at source) & transmission & (at sample) &  W/cm$^2$ \\
        \hline \hline
        14.3 & 7.0 & 45 & 0.10 & 4.5 & $2.3 \times 10^{14}$  \\
        15.9 & 10.2 & 95 & 0.13 & 12.4 &  $6.2 \times 10^{14}$\\
        19.1 &  16.6 & 84  & 0.23 & 19.3 & $ 9.5 \times 10^{14}$ \\
        \hline
    \end{tabular*}
\end{table}
The estimate of the pulse energy at $\hbar \omega$=14.3 eV was indirect, since the FERMI intensity monitors do not function at this energy, because they are based on ionization of nitrogen gas, and the photon energy is below the threshold for ionization. The method employed was to first use the in-line spectrometer to measure spectra at 15.9 eV energy and simultaneously the pulse energies from the gas cell monitors, which gave a calibration of the spectrometer intensity versus pulse energy at this wavelength. Then spectrometer spectra were measured at 14.3 eV, and corrected for grating efficiency and detector sensitivity, to yield pulse energies.

\section{\label{sec:theory}Theory}
We now consider the physics of the experiment from two theoretical points of view:  real-time {\it ab initio} simulations, which are very accurate, but computationally expensive; and
perturbation theory, which allows us to 
explore the physics analytically and gain insights with relatively low computational costs.

\subsection{Real-time {\it ab initio} simulations}

We numerically computed the photoionization of Ne irradiated by two-color XUV pulses, using the time-dependent complete-active-space self-consistent field (TD-CASSCF) method 
\cite{Sato2013,Sato2016}, and the parameters in Table \ref{tab:Table2}. 
The pulse length was chosen to be 10 fs for reasons of computational economy. It has been shown that the pulse length does not affect the result, provided the photoionization is non-resonant, i.e. no resonances occur within the photon bandwidth \cite{Ishikawa2012, Ishikawa2013}. As a further check, we also calculated the phase shift difference at 14.3 eV photon energy for pulse durations of 5, 10 and 20 fs, and found identical results.
Thus we can safely scale the results to the present longer experimental pulses.

Neither the absolute intensity nor the ratio of intensities of the harmonics influences the calculated phase, as we show below.
The dynamics of the laser-driven multielectron system is described by the time-dependent Schr\"odinger equation (TDSE)\index{Time-dependent Schr\"odinger equation (TDSE)}:
\begin{equation}
\label{eq:TDSE}
i\frac{\partial\Psi (t)}{\partial t} = \hat{H}(t)\Psi (t),
\end{equation}
where the time-dependent Hamiltonian is
\begin{equation}
\hat{H}(t)=\hat{H}_1(t)+\hat{H}_2,
\end{equation}
with the one-electron part
\begin{equation}
\label{eq:H1}
\hat{H}_1(t) = \sum_{i=1}^{Z} \hat{h}({\bf r}_i,t) 
\end{equation}
and the two-electron part
\begin{equation}
\hat{H}_2 = \sum_{i=1}^{Z} \sum_{j < i} \frac{1}{|{\bf r}_i - {\bf r}_j|}.
\end{equation}

We employ the velocity gauge for the laser-electron interaction in the one-body Hamiltonian:
\begin{equation}
\label{eq:velocity-gauge}
\hat{h}({\bf r},t) = \frac{{\bf k}^2}{2} + {\bf A}(t) \cdot {\bf k} -\frac{Z}{|{\bf r}|},
\end{equation}
where ${\bf A}(t) = -\int {\bf E}(t)dt$ is the vector potential,
and ${\bf E}(t)$ is the laser electric field,
see Eq.~(\ref{eq:Efield}), and $Z$ (=10 for Ne) the atomic number.

\begin{table}[h]
\caption{\label{tab:Table2}Parameters used for theoretical calculations. $I_\omega$ and $I_{2\omega}$ are the peak intensities of the fundamental and second harmonic respectively.
}
    \centering
    \begin{tabular*}{\textwidth}{@{}l*{15}{@{\extracolsep{0pt plus12pt}}l}}
        \hline
        $\hbar \omega$, eV & $I_\omega$, W/cm$^2$   & $I_{2\omega}$, W/cm$^2$   \\
        \hline \hline
        14.3 & $1 \times 10^{13}$ & $2.32 \times 10^{8}$    \\
        15.9 & $1 \times 10^{13}$ & $1.18 \times 10^{8}$  \\
        19.1 &  $1 \times 10^{13}$ & $2.82 \times 10^{8}$   \\
        \hline
    \end{tabular*}
\end{table}

In the TD-CASSCF method, the total electronic wave function is given in the configuration interaction (CI) expansion:
%%%%%%
\begin{gather}
	\Psi ({x}_1, {x}_2, \cdots, {x}_N, t) = \sum_{I} C_I(t) \Phi_I({x}_1, {x}_2, \cdots, {x}_N, t).
\end{gather}
where $x_n = \{ {\bf r}_n, \sigma_n \}$ is the joint designation for spatial and spin coordinates of the $n$-th electron.
The electronic configuration $\Phi_I({x}_1, {x}_2, \cdots, {x}_N, t)$ is a Slater determinant composed of spin orbital functions $\{\psi_p({\bf r},t) \times s(\sigma)\}$, where $\{ \psi_p({\bf r},t) \}$ and $\{ s(\sigma) \}$ denote spatial orbitals and spin functions, respectively. 
Both the CI coefficients $\{C_I\}$ and orbitals vary in time.

The TD-CASSCF method classifies the spatial orbitals into three groups: doubly occupied and time-independent frozen core (FC), doubly occupied and time-dependent dynamical core (DC), and fully correlated active orbitals:
\begin{equation}
	\Psi = \hat{A}\left[\Phi_{\rm fc}\Phi_{\rm dc}\sum_I\Phi_I C_I\right],
\end{equation}
where $\hat{A}$ denotes the antisymmetrization operator, $\Phi_{\rm fc}$ and $\Phi_{\rm dc}$ the closed-shell determinants formed with numbers $n_{\rm fc}$ FC orbitals and $n_{\rm dc}$ DC orbitals, respectively, and $\{\Phi_I\}$ the determinants constructed from  $n_{\rm a}$ active orbitals. 
We consider all the possible distributions of active electrons among  active orbitals.
Thanks to this decomposition, we can significantly reduce the computational cost without sacrificing the accuracy in the description of correlated multielectron dynamics.
The equations of motion that describe the temporal evolution of the CI coefficients $\{C_{I}\}$ and the orbital functions $\{\psi_p\}$ are derived by use of the time-dependent variational principle \cite{Sato2013}. The numerical implementation of the TD-CASSCF method for atoms is detailed in Refs.~\cite{Sato2016,Orimo2018}.

\subsection{Extraction of the photoelectron angular distribution and the phase shift difference}

From the obtained time-dependent wave functions, we extract the angle-resolved photoelectron energy spectrum (ARPES) by use of the time-dependent surface flux (tSURFF) method \cite{Tao2012}. 
This method computes the ARPES from the electron flux through a surface located at a certain radius $R_s$, beyond which the outgoing flux is absorbed by the infinite-range exterior complex scaling \cite{Scrinzi2010,Orimo2018}.

We introduce the time-dependent momentum amplitude $a_p(\boldsymbol{k},t)$ of orbital $p$ for photoelectron momentum ${\bf k}$, defined by
\begin{gather}
	a_p({\bf k},t) =  \ME{\chi_{\bf k}({\bf r},t)}{u(R_s)}{\psi_p ({\bf r},t)}
    \equiv \int_{r>R_s}\chi_{{\bf k}}^*({\bf r},t)\psi_p({\bf r},t)d^3 {\bf r},
\end{gather}
where $\chi_{{\bf k}}({\bf r},t)$ denotes the Volkov wavefunction, and $u(R_s)$ the Heaviside function which is unity for $r>R_s$ and vanishes otherwise. The use of the Volkov wavefunction implies that we neglect the effects of the Coulomb force from the nucleus and the other electrons on the photoelectron dynamics outside $R_s$, which has been confirmed to be a good approximation \cite{Orimo2018b}.
The photoelectron momentum distribution $\rho ({\bf k})$ is given by
\begin{equation}
	\label{eq:PEMD}
	\rho({\bf k}) 
	=  \sum_{pq} a_p({\bf k},\infty) a_q^*({\bf k},\infty) \ME{\Psi(t)}{\hat{E}^q_p }{\Psi(t)},
\end{equation}
with $\hat{E}^q_p\equiv \sum_\sigma \hat{a}_{q\sigma}^\dagger \hat{a}_{p\sigma}$.
One obtains $a_p({\bf k},\infty)$ by numerically integrating:
\begin{equation}
	\label{eomme_tsurff}
	-i \frac{\partial}{\partial t} a_p({\bf k},t) = \ME{\chi_{{\bf k}}(t)}{[\hat{h}_s, u(R_{s})]}{\psi_p(t)} 
	+ \sum_q a_q({\bf k},t) \{\ME{\psi_q(t)}{\hat{F}}{\psi_p(t)} - R^q_p\},
\end{equation}
where $\hat{h}_s=\frac{{\bf k}^2}{2}+{\bf A}(t)\cdot {\bf k}$, $R^q_p=i\Skalar{\psi_q}{\dot{\psi}_p} - \ME{\psi_q}{\hat{h}}{\psi_p}$, and $\hat{F}$ denotes a nonlocal operator describing the contribution from the inter-electronic Coulomb interaction \cite{Sato2016,Orimo2018}.
The numerical implementation of tSURFF to TD-CASSCF is detailed in Ref.~[\onlinecite{Orimo2018b}].

We evaluate the photoelectron angular distribution $I \left( \theta;\phi \right)$ as a slice of $\rho \left( {\bf k} \right)$
at the value of $\left| {\bf k} \right|$ corresponding to the photoelectron peak, and as a function of the optical phase $\phi$.
Then, employing a fitting procedure very similar to that used for the experimental data, we extract the phase shift difference $ \Delta \tilde \eta$ between single-photon and two-photon ionization at photoelectron energies 7.0~eV, 10.2~eV and 16.6~eV. The results are shown in Fig.~\ref{fig:2}. 

\subsection{\label{sec:TheoPT}Perturbation theory}
In the experiment, the number of optical cycles in the pulse is of the order of 400 for the fundamental
and therefore we can treat the field as having constant amplitude and omit
the initial phase of the field with respect to the envelope (carrier-envelope phase).
Within the perturbation theory, we checked that our final results with
an envelope including 100 optical cycles or more differ only within the optical linewidth
from those obtained with the constant amplitude field.
The bichromatic electric field is then described by Eq.~(\ref{eq:Efield}), with time-independent $I_{\omega}$ and $I_{2\omega}$.
 The calculations described below were carried out for 384 optical cycles and a peak intensity of 
 $1 \times 10^{12}$ W/cm$^2$. However neither the absolute intensity nor the ratio of intensities of the harmonics influences the calculated phase, as we show below.

We make two main assumptions: the dipole approximation for the interaction of the atom with the classically
described electromagnetic field, and the validity of the lowest nonvanishing order perturbation theory with respect
to this interaction.
These approximations are well fulfilled for neon in the FEL spectral range and intensities of interest here.
We expand the amplitudes in the lowest non-vanishing order of perturbation 
theory in terms of matrix elements of the operator of evolution \cite{Messiah1961}. The expansion implies that in the second-order amplitude all virtual intermediate states are taken into account. Excitations of the seven lowest and most important intermediate dipole-allowed states originating from configurations 
($2p^5$ $3s,4s,3d$) were accounted for accurately within the multiconfiguration
intermediate-coupling approximation with relativistic Breit-Pauli corrections in the atomic Hamiltonian. 
All other virtual states (of infinite number), including those in the continuum, were accounted for by a variationally stable method \cite{Gao1989, Staroselskaya2015} in the Hartree-Fock-Slater approximation.
More details can be found in \cite{Gryzlova2018}. Further derivations within the independent particle approximation
are given in the Appendix.

\section{\label{sec:Results}Results}
\begin{figure}[bt]
    \centering
    \includegraphics{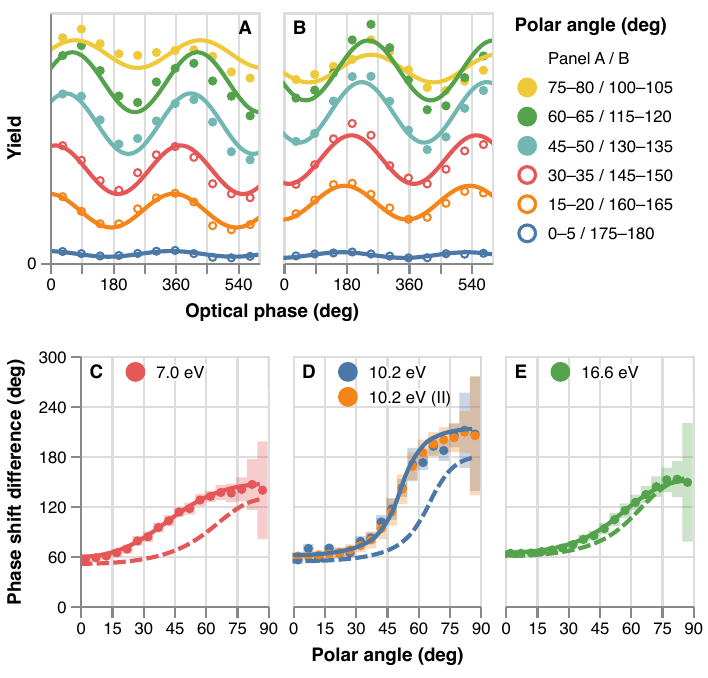}
    \caption{~\label{fig:2}Upper panels: typical photoelectron yields $I \left( \theta; \phi \right)$
as a function of optical phase $\phi$
at intervals of polar angles $\theta$.  
The signal was integrated over the $5^{\circ}$ intervals shown on the right. The curves have not been shifted, thus their vertical position reflects directly the value $A_0$ in Eq.~(\ref{eq:yield}).
The photoelectron kinetic energy is 7.0~eV.
Circles are experimental results; 
lines are sinusoidal fits of the experimental results.
Lower panels: extracted phase shift differences as a function of the polar angles, for four data sets and three photoelectron kinetic energies:
left (C): 7.0~eV; middle (D): 10.2~eV; right (E): 16.6 eV.
Circles are experimental results; shaded areas show their uncertainties.
Dashed lines: perturbation theory; solid lines: real time \textit{ab initio} theory. Note that the curves in panels A and B oscillate in antiphase, because they correspond to emission directions on opposite sides of the photon propagation direction.}
\end{figure}

We extracted  $\Delta \tilde \eta \left( \theta \right)$ from the measured PADs
at three combinations of $\omega$ and $2\omega$ (corresponding to photoelectron kinetic energies,
7.0~eV, 10.2~eV and 16.6~eV), at each $5^{\circ}$ interval of
polar angle. 
The spatial and temporal symmetry properties of the system impose constraints on the oscillatory behavior of the two emission hemispheres. Upon reflection in a plane perpendicular to the electric vector ($\theta \rightarrow \pi - \theta$), the electric field defined in Eq.~(\ref{eq:Efield}) is inverted: 
$E\left( t \right) \rightarrow -E\left( t \right)$,
and the $\omega$-$2\omega$ relative phase becomes $\phi + \pi$.
From the arguments above, Eq.~(\ref{eq:mod}) becomes
\begin{equation}
    \label{eq:yield2}
     I (\pi - \theta, \phi + \pi) = \;
        A_0 + A \cos \left[ \phi + \pi -  \Delta \tilde \eta \left( \pi - \theta \right) \right]
        =  A_0 - A \cos \left[ \phi -  \Delta \tilde \eta \left( \pi - \theta \right) \right],
\end{equation}
where we have omitted the argument $\bar{\epsilon}$ and included explicitly the argument $\theta$. Comparison with Eq.~(\ref{eq:yield}) indicates that the intensities at the two opposite angles oscillate in antiphase, that is, 
$ \Delta \tilde \eta \left( \pi-\theta \right) =  \Delta \tilde \eta \left( \theta \right) + \pi$. It can be seen in Figs.~\ref{fig:2}A--\ref{fig:2}B that the experimental data does indeed oscillate in antiphase for each angular interval over $\theta$. Since this is a symmetry constraint, it was imposed in the analysis of the data.

In Figs.~\ref{fig:2}C--~\ref{fig:2}E, it can be seen that there is a significant increase of $ \Delta \tilde \eta \left( \theta \right)$ at $\sim 90^{\circ}$, 
especially for 10.2~eV. The angular dependent variations of $ \Delta \tilde \eta \left( \theta \right)$ 
at 7.0~eV and 16.6~eV are similar. 
We performed calculations for the phase shift differences 
$ \Delta \tilde \eta \left( \theta \right)$ using both  perturbation theory and
real time \textit{ab initio} methods (see Theory Sec. \ref{sec:theory} and Appendix A) and both theories reproduce well the observed behavior, see {Figs.~\ref{fig:2}C--~\ref{fig:2}E}.
The perturbation theory result at the intermediate angles is very sensitive to the contribution of the p-d-f two-photon ionization path, which may not be accurately reproduced by the local-potential approximation in summation over the Rydberg and continuum $d$-states.

\begin{figure}[bt]
    \centering
\includegraphics[width=3 in]{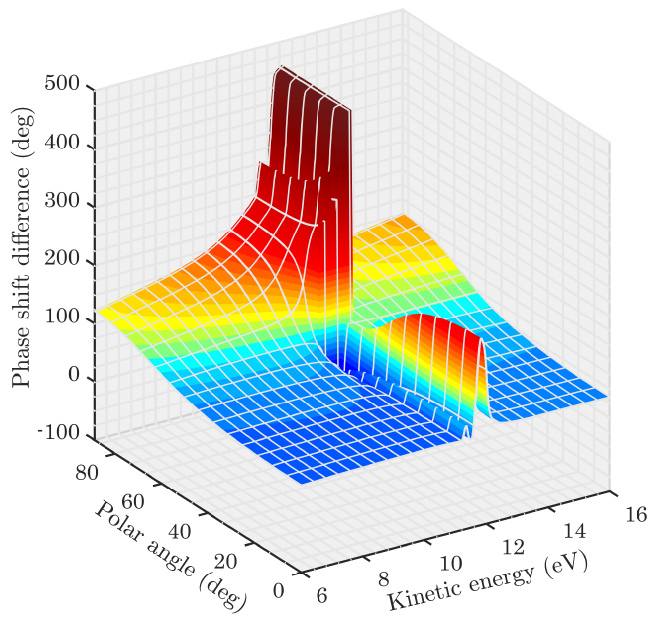}
    \caption{~\label{fig:PT_phase} Phase shift differences $ \Delta \tilde \eta$ %(left) right:  
    of two-photon ionization relative to single-photon ionization, calculated by perturbation theory,
as a function of the polar angle $\theta$
and photoelectron kinetic energy.
The large variation near 12.0~eV kinetic energy is due to the  $2p \rightarrow 3s$ resonance.
}
\end{figure}

Figure~\ref{fig:PT_phase} shows the theoretical dependence of $ \Delta \tilde \eta \left( \theta \right)$
on electron kinetic energy and polar angle $\theta$, calculated using perturbation theory.
There is a single-photon $2p \rightarrow 3s$ resonance of the fundamental wavelength at 16.7~eV photon energy (12~eV kinetic energy for the two-photon/second harmonic). The behavior of $ \Delta \tilde \eta$ in the region of the resonance is complicated:
we can clearly see that $ \Delta \tilde \eta \left( \theta \right)$ at $\theta \sim 90^{\circ}$ increases near the resonance around 
12~eV
and then returns to a value similar to that at $\sim 7$~eV.
This indicates that the large phase shift difference observed at 10.2~eV in {Fig.~\ref{fig:2}D}
is due to the influence of the resonance at 12.0~eV~\cite{Banerjee2019,Cirelli2018}, and suggests that future experiments should explore this region in fine detail, to observe the predicted rapid changes in $ \Delta \tilde \eta$.
Both theories reproduce this behavior well, with
the time-dependent \textit{ab initio} method exhibiting excellent agreement,
validating the present experimental method.

We show in Appendix A that the method is independent of the relative intensities of the fundamental and second harmonic radiation, see Eqs. (A5) to (A9). This is a considerable advantage from an experimental point of view, as it is not necessary to measure precisely the intensity and focal spot shape. Furthermore, there are no effects due to volume averaging over the Gaussian spot profile, or over the duration of the pulses. We verified this experimentally for the kinetic energy of 16.6 eV, Fig. 2D, where the ratio of ionization rates was 1:4 (rather than 1:2 used for the other energies), and the experiment and theory agree well. 
 
%***************************************************
\section{\label{sec:Discusson}Discussion}

In this section we elucidate the relationship of our data, i.e. photoelectron angular distributions created by collinearly polarized biharmonics, to time-delay studies 
described in the introduction. 
We limit ourselves to the case where any discrete state in the continuum (autoionizing state) or in the discrete spectrum lies outside the bandwidth of the pulses ($<$0.02 eV in the present case).
These conditions are well fulfilled in our experiments. 
The resonant case has been discussed elsewhere~\cite{Argenti2017}.
 
We first consider the simple situation of photoionization from a spherically symmetric orbital $s$. The present method can be extended straightforwardly to
inner shell ionization of atoms, such as $1s^2$ of Ne.  
Single-photon ionization leads to a continuum state with angular momentum $p$, while two-photon ionization leads to two final quantum states $s$ and $d$. 
Then the PAD $I_e(\theta)$ is described by
\begin{align}
\label{eq:PAD1a}
I_e(\theta) & = \left| c_s e^{i\eta_{s}}Y_{00}(\theta,\varphi) + c_pe^{i(\eta_{p}+\phi)}Y_{10}(\theta,\varphi) + c_d e^{i\eta_{d}}  Y_{20}(\theta,\varphi)\right|^2   \,,
\end{align}
where $c_s$, $c_p$, and $c_d$ are real-valued partial-wave amplitudes and $\eta_s$, $\eta_p$, and $\eta_d$ are the corresponding arguments.
$I_e(\theta)$ can also be expressed as
\begin{equation}
\label{eq:PAD2a}
I_e(\theta) = \left(c_s^2 + c_p^2 + c_d^2\right)
\left[1+\sum_{l=1}^{4}\beta_l P_l(\cos\theta)\right] \,,  
\end{equation}
where $\ P_l(\cos\theta)$ are the Legendre polynomials describing the angular distributions and $\beta_l$ are the corresponding asymmetry parameters. 
After some algebra, we have~\cite{DiFraia2019}
\begin{eqnarray}
\label{eq:beta3sa}
\beta_3
& = & \frac{
	6 \sqrt{15} c_{d} c_{p} \cos{\left (- \eta_{d} + \eta_{p} + \phi \right )}
}{5 \left( c_{d}^{2} + {c_{p}}^{2} + c_{s}^{2} \right)}
\equiv \left[\beta_3\right]_0 \cos \left( \phi  - (\eta_d-\eta_p) \right) \,, \\
\label{eq:beta1-(2/3)beta3a}
\beta_1 - \frac{2}{3}\beta_3             
& =  & \frac{
	2 \sqrt{3} c_{p} c_{s} \cos{\left (\eta_{p} - \eta_{s} + \phi \right )}
}{c_{d}^{2} + {c_{p}}^{2} + c_{s}^{2}}
\equiv \left[\beta_1 - \frac{2}{3}\beta_3\right]_0 \cos{\left( \phi-(\eta_s-\eta_p) 
\right) } \;,              
\end{eqnarray}
where $[\beta_3]_0$ and $[\beta_1 - \frac{2}{3}\beta_3]_0$ are constants. 
Thus, if we record PADs as a function of $\phi$ and extract $\beta_l$ ($l=1$ to 4), 
we can directly read off $\eta_d-\eta_p$ and 
$\eta_s-\eta_p$ from the oscillations 
of $\beta_3$ and $\beta_1 - \frac{2}{3}\beta_3$ using  Eqs.~(\ref{eq:beta3sa}) 
and~(\ref{eq:beta1-(2/3)beta3a}).
Let us recall that the Wigner delay of each partial wave, $\tau_l$,
corresponds to the energy derivative of the argument of the amplitude
(note that $Y_{l0}(\theta, \varphi)$ are real),
$\tau_l(\epsilon) = \frac{{d\eta}_l(\epsilon)}{d\epsilon}$~\cite{Wigner1955}. 
By measuring $\eta_d-\eta_p$ and $\eta_s-\eta_p$ as a function of energy, one can take the energy derivative and obtain the Wigner delay differences $\tau_l(\epsilon)-\tau_p(\epsilon)$ with $l =s$ and $d$.
In simple models, like the Hartree-Fock approximation,  
$\frac{{d\eta}_l(\epsilon)}{d\epsilon} = \frac{{d\delta}_l(\epsilon)}{d\epsilon}$,
where $\delta_l(\epsilon)$ is the scattering phase, while in more complicated
cases, an extra energy dependent phase may be acquired by the partial amplitude~\cite{Watzel2015}.

We now group the $s$ and $d$ waves as a two-photon-ionization wave packet. Then the photoelectron wave packet in a given direction $\theta$ sufficiently far from the nucleus, and the corresponding PAD are expressed as Eqs.~\eqref{eq:wavepacket2} and \eqref{eq:yield}, respectively.
The energy derivative of  $\Delta \eta \left( \theta \right) \equiv \eta_{\omega}(\theta) - \eta_{2\omega}(\theta)$
is a difference between the group delays of the two wave packets, generated by two-
and single-photon ionization, respectively. 
In the original photoemission delay experiment \cite{Schultze2010} with attosecond streaking, for example, Ne $2s$ and $2p$ electrons were ionized by an attosecond pulse to different final kinetic energies. As a result, the more energetic photoelectron from $2p$ arrived at the detector much earlier than that from $2s$, regardless of the measured delay. The situation is similar for subsequent measurements using streaking and RABBITT.
In great contrast, in the present case, both single- and two-photon ionization result in the same photoelectron energy. Therefore, the single- and two-photon-ionization wave packets actually reach a given distance with a relative (group) delay given by $\frac{\partial\Delta\eta}{\partial\epsilon}$.

By comparing Eqs.~\eqref{eq:yield} and \eqref{eq:PAD1a}, we can describe the phase factor $e^{i\Delta \eta \left( \theta \right)}$ with $\Delta \eta \left( \theta \right) \equiv \eta_{\omega}(\theta) - \eta_{2\omega}(\theta)$ being the angle-resolved phase difference between the two-photon and single-photon ionization amplitudes as,
\begin{align}
\label{eq:phase-exponential-form}
	A(\theta)e^{i\Delta\eta(\theta)}
	= 2c_pY_{10}(\theta, \varphi)
	\left[
	c_d Y_{20}(\theta, \varphi) \, e^{i(\eta_d-\eta_p)}
	+ c_sY_{00}(\theta) \, e^{i(\eta_s-\eta_p)}
	\right].
\end{align}
Thus, the phase factor $e^{i\Delta \eta \left( \theta \right)}$ is the coherent (i.e., with respect to amplitudes) average of $e^{i(\eta_{d} - \eta_{p})}$ ($d$-$p$ interference) and $e^{i(\eta_{s}  - \eta_{p})}$ ($s$-$p$ interference) with the relative weight,
\begin{equation}
W (\theta) = B \, g(\theta),\qquad B= \frac{c_d}{c_s}, \qquad g(\theta) = \frac{Y_{20}(\theta, \varphi)}{Y_{00}(\theta, \varphi)} =
\frac{\sqrt{5}}{4} \left[ 3 \cos (2\theta) + 1 \right] \,.
\end{equation}
In other words, $\Delta \eta \left( \theta \right)$ can be regarded as a $vectorial$ average 
of $\eta_s - \eta_p$ and $\eta_d - \eta_p$ with the relative weight $W(\theta)$.
Equivalently, $\Delta \eta (\theta)$ may be presented as
\begin{align}
\label{eq:Phase}
\tan\Delta \eta \left( \theta \right) 
=  \frac{\cos (\eta_s - \eta_p) + W(\theta) \, \cos (\eta_d - \eta_p)}{\sin(\eta_s - \eta_p) + W(\theta) \, \sin (\eta_d - \eta_p)} \,.  
\end{align}
The energy derivative of $\Delta \eta (\theta)$ does not give us additional information
about the photoionization amplitudes, but provides us with the group delay and
may enhance the sensitivity to the energy-dependent behavior of the two-photon ionization amplitudes, as described below.

Note two important characteristics of $\Delta \eta \left( \theta \right)$: 
(i) $\Delta \eta \left( \theta \right)$ exhibits a quasi-cosine shape, and monotonic
dependence on $\theta$ due to the geometric factor $g(\theta)$ (see Fig.~\ref{fig:2}(c)-(e) and Appendix A) and (ii) 
$\Delta \eta \left( \theta \right)$ is sensitive to the two-photon ionization dynamics due to the dynamical factor $B$. For example, 
if the two-photon pathways are close to an intermediate discrete resonance (but still well outside the bandwidth), the group delay difference $\frac{\partial\Delta\eta}{\partial\epsilon}(\theta)$ is sensitive to it through rapid change in $B$, while  $\frac{d\eta_s}{d\epsilon},\frac{d\eta_p}{d\epsilon}$, and $\frac{d\eta_d}{d\epsilon}$ are small individually, as can be seen in Fig.~\ref{fig:PT_phase}.

We now turn to photoionization from a $p$ orbital, which includes the present case of Ne 2$p$ ionization, and is more complicated.
The complexity arises from two sources. We have three incoherent contributions from $m=0$
and $\pm 1$ for the magnetic sublevels of the remaining ion core Ne$^+$ and four
contributions of partial waves, $s$, $p$, $d$, and $f$ in the photoelectron wavepacket. 
Detailed derivations of the equations describing the PADs are given in Appendix A and here we describe only the results 
relevant to the present discussion. 
For $m=\pm 1$, we have a pair of two-photon pathways via $d$ intermediate states, i.e., 
$p \rightarrow d \rightarrow p$ and $p \rightarrow d \rightarrow f$, together with
the single-photon pathway $p \rightarrow d$. Thus, only three partial waves are involved 
and therefore a discussion similar to that above for $s$ ionization holds. 

For $m=0$, we have two-photon pathways via $s$ and $d$ intermediate states, leading 
to $p$ and $f$ final states, together with single-photon ionization that leads to $s$ 
and $d$ final states. Thus, there are four partial waves involved. 
Although we can derive equations similar to Eqs.~(\ref{eq:beta3sa}) and
~(\ref{eq:beta1-(2/3)beta3a}) (see Eqs.~(\ref{eq:beta1})-(\ref{eq:beta4})), what we 
can extract from the measurement is only a vectorial average of phase differences
$\eta_l - \eta_{l'}$ between even and odd different partial waves $l$, $l'$. 
We can define the angle-resolved phase difference $\Delta \eta_m(\theta)$ for each $m$
(see Eq.~(\ref{eq:mod})), which is also a vectorial average of $\eta_l - \eta_{l'}$. 
Similar to ionization from the $s$ state,
the energy derivative of $\Delta \eta_m(\theta)$ may be regarded as an angle-resolved
group delay between single- and two-photon wave packets for each $m$.%, 

In the experiment, we measured an 
(incoherently) weighted average $\Delta \tilde\eta$ of angle-resolved
phase differences $\Delta\eta_m$ of different $m$ as defined in Eqs.~(\ref{eq:mod}) and~(\ref{eq:summation}). 
One can introduce the energy derivative of the weighted average phase difference 
$\Delta \tilde\eta(\theta)$,  
and may call it generalized delay, but this definition of time delay is 
different from that commonly employed for the time delay of an incoherent sum of 
wavepackets. Usually the phase of each wavepacket is first differentiated with respect to energy and then averaged over $m$ \cite{Smith1960}, while in this study,
 $\frac{d\Delta \tilde{\eta}}{d\epsilon}(\theta)$  
first averages the wavepacket phase over $m$ and then differentiates it with respect to 
the photoelectron energy. 
\\

%*****************************************************

\section{\label{sec:summary}Summary and outlook}

In this work we have described a new method to determine angle-resolved relative phase between single- 
and two-photon ionization amplitudes, 
and used it to measure the $2p$ photoionization of Ne. 
Our approach allows us to explore the phase difference between different ionization pathways, e.g., those of odd and even parities, with the same photoelectron energy. 

The method is based on FEL radiation, so that it can be extended to shorter wavelengths, eventually to inner shells,  
which lie in a wavelength region where optical lasers have reduced pulse energy. 
This is an important addition to the armoury of techniques available to attosecond science 
and gives access to the 
phase difference between single- 
(odd parity) and two-photon (even parity) transition amplitudes, or the energy variation of the phase of two-photon ionization amplitudes affected by the intermediate resonances, as seen in the Ne $2p$ photoionization.
For $ns^2$ subshells of atoms, e.g., $1s^2$ of He, $1s^2$ and $2s^2$ of Ne, etc. 
in particular, 
one can extract the eigenphase differences for $s$, $p$, and $d$ partial waves of electron-ion scattering, and their energy derivatives correspond to the Wigner delay difference of the partial waves. 
This method is also applicable to molecules.

While it does not yet appear to be feasible with present HHG sources, it may become possible in the future, but there are many technical challenges. Since HHG sources produce a frequency comb, the chief technical challenges are to filter the beam to achieve bichromatic spectral purity, maintain attosecond temporal resolution, and provide enough pulse energy at the fundamental wavelength to initiate two-photon ionization. Furthermore, HHG sources have not yet demonstrated the level of phase control which we have at our disposal. Given the rapid progress in HHG sources, these conditions may eventually be met, in which case our method will become more widely accessible.

The information obtained by this method is complementary to that of streaking and RABBITT methods, in the sense that different phase differences are measured. 
We have directly measured the angle-resolved average phase difference $\Delta \tilde \eta \left( \theta \right)$ of two-photon amplitude relative to the single-photon ionization amplitude. 
The basic physics giving rise to its angular dependence is related to interference between photoelectron waves 
emitted in one- and two-photon ionization, consisting of partial photoelectron waves with opposite parities. We have shown that the overall shape of $\Delta \tilde \eta \left( \theta \right)$ versus angle can be understood qualitatively.

\begin{acknowledgments}
This work was supported in part by  
the X-ray Free Electron Laser Utilization Research Project 
and the X-ray Free Electron Laser Priority Strategy Program of the Ministry of Education, Culture, Sports, Science, and Technology of Japan (MEXT)  
and the IMRAM program of Tohoku University,  
and the Dynamic Alliance for Open Innovation Bridging Human, Environment and Materials program.  
K.L.I. gratefully acknowledges support by the Cooperative
Research Program of the Network Joint Research Center
for Materials and Devices (Japan), Grant-in-Aid for
Scientific Research (Grants No. 16H03881, No. 17K05070, No. 18H03891, and No. 19H00869) from MEXT, 
JST COI (Grant No.~JPMJCE1313), JST CREST (Grant No.~JPMJCR15N1), MEXT Quantum Leap Flagship Program (MEXT Q-LEAP) Grant No.~JPMXS0118067246, and Japan-Hungary Research Cooperative Program, JSPS and HAS.
E.V.G. acknowledges the Foundation for the Advancement of Theoretical Physics and Mathematics ``BASIS."
D.Y. acknowledges supports by
JSPS KAKENHI Grant Number JP19J12870,
and a Grant-in-Aid of Tohoku University Institute
    for Promoting Graduate Degree Programs Division for Interdisciplinary
    Advanced Research and Education.
We acknowledge the support of the Alexander von
Humboldt Foundation (Project Tirinto), the Italian Ministry
of Research Project FIRB No. RBID08CRXK and No. PRIN
2010 ERFKXL 006, the bilateral project CNR JSPS
Ultrafast science with extreme ultraviolet Free Electron
Lasers, and funding from the European Union Horizon
2020 research and innovation program under the Marie
Sklodowska-Curie Grant Agreement No. 641789 MEDEA
(Molecular Electron Dynamics investigated by IntensE Fields
and Attosecond Pulses), and
the RFBR and DFG for the research project No. 20-52-12023.
T.M. and M.M. acknowledge support by Deutsche Forschungsgemeinschaft Grant No. SFB925/A1.
We thank the machine physicists of
FERMI for making this experiment possible by their excellent
work in providing high quality FEL light.
\end{acknowledgments}

\appendix
\section{\label{sec:Appendix}Perturbation theory. Derivation of equations in the independent particle model}

In addition to the approximations described in Section \ref{sec:TheoPT} (the dipole approximation, the validity of the lowest nonvanishing order perturbation theory), here we add the LS-coupling approximation within the independent particle model. The photoelectron angular distribution $I(\theta; \phi)$ of a Ne 2$p$ electron can be derived
by standard methods~\cite{Amusia1990} in the form
\begin{equation} \label{eq:apad1}
I(\theta; \phi)= I_0 \sum_{m=0, \pm 1} \left| \sum_{\xi} C^m_{\xi}(\phi) Y_{\ell_{\xi} m}(\theta, \varphi) \right|^2 \,,
\end{equation}
where $m$ is the magnetic quantum number of the initial $2p$ electron, $Y_{\ell m}(\theta, \varphi)$ is 
a spherical harmonic in the Condon-Shortley phase convention, $I_0$ is a normalization factor irrelevant to further discussion; note that the dependence on $\varphi$ cancels out. 
The complex coefficients $C^m_{\xi}(\phi)$ depend on ionization amplitudes,
 and the index $\xi$ denotes the ionization path. For single-photon  
ionization $\xi = \ell_{\xi}$, where $\ell_{\xi}$ is the orbital momentum of the photoelectron 
with possible values $\ell_{\xi} = s, \, d$. For two-photon ionization $\xi = \{\ell_{\xi}, \ell'_{\xi} \}$, 
where $\ell'_{\xi}$
is the orbital momentum of the virtual intermediate state, with possible combinations $\xi = ps, \, pd, \, fd$.

After applying the Wigner-Eckart theorem~\cite{Varshalovich1988} to factor out the dependence on the projection $m$, 
the coefficients $C^m_{\xi}(\phi)$ may be expressed as (for brevity, we omit the argument $\phi$ when writing the coefficients):
\begin{equation} \label{eq:amat1}
C_s^0  =  - \frac{1}{\sqrt{3}} D_s \, e^{i \phi}, \;\; C_d^0  =  \sqrt{\frac{2}{15}} D_d \,e^{i \phi},
\;\;  C_d^{\pm 1}  =  \frac{1}{\sqrt{10}} D_d \, e^{i \phi},
\end{equation}
\begin{eqnarray} \label{eq:amat2}
C^0_{ps} = -\frac{1}{3} D_{ps}, \;\; C^{\pm 1}_{ps} = 0, \nonumber\\
C^0_{pd}  = - \frac{2}{15} D_{pd}, \;\; C^{\pm1}_{pd} = - \frac{1}{10} D_{pd}, \nonumber\\
C^0_{fd} = \frac{\sqrt{2}}{5\sqrt{7}} D_{fd}, \;\; C^{\pm1}_{fd} = \frac{2}{5 \sqrt{21}} D_{fd}.
\end{eqnarray}
Here 
\begin{equation} \label{eq:ddn}
D_{\xi} = d_{\xi} \, e^{i \eta_{\xi}}
\end{equation}
are complex reduced matrix elements, independent of $m$, with magnitude 
$d_{\xi} = \left| D_{\xi} \right|$ and phase $\eta_{\xi}$. 
Note that one- (first order) and two-photon (second order)
matrix elements~(\ref{eq:ddn}), both marked by a single index $\xi$, 
are respectively proportional to the square root of intensity, and to intensity, of the associated field.

Equation~(\ref{eq:apad1}) can be readily cast into the form~(\ref{eq:yield}), where
\begin{eqnarray} \label{eq:a0}
A_0 & = & \frac{I_0}{4 \pi} \sum_{\lambda = 0,2,4} Z_{\lambda} P_{\lambda}(\cos \theta) , \\
A &  =  & \frac{I_0}{4 \pi} \, N  \,, \label{eq:aa} \\
\cos \Delta \tilde \eta  & = & N^{-1} \sum_{\lambda = 1,3} {\rm Re} \, Z'_{\lambda} P_{\lambda} (\cos \theta) \,, \label{eq:cos} \\
\sin \Delta \tilde \eta  & = & N^{-1} \sum_{\lambda = 1,3} {\rm Im} \, Z'_{\lambda} P_{\lambda} (\cos \theta) \,, \label{eq:sin} \\
N & = & \left| \sum_{\lambda= 1,3} Z'_{\lambda} P_{\lambda} (\cos \theta) \right| \,, \label{eq:n}
\end{eqnarray}
\begin{equation} \label{eq:z}
    Z_{\lambda} = \sum_{\substack{m=0, \pm1\\ \xi=s,d,ps,pd,fd\\ \bar{\xi}=s,d,ps,pd,fd}} (-1)^m C_{\xi}^m C_{\bar{\xi}}^{m \ast} 
    \sqrt{(2\ell_{\xi}+1)(2\ell_{\bar{\xi}}+1)}
   \CGC{\ell_{\xi}}{m}{\ell_{\bar{\xi}}}{-m}{\lambda}{0} 
   \CGC{\ell_{\xi}}{0}{\ell_{\bar{\xi}}}{0}{\lambda}{0} \,,
\end{equation}
where $\CGC{j_1}{m_1}{j_2}{m_2}{j}{m}$ are Clebsch-Gordan coefficients~\cite{Varshalovich1988} and   $Z'_{\lambda} = Z_{\lambda}|_{\phi=0}$.
In particular
\begin{equation}
	Z_0 = \frac{1}{30} \Big[
       10 (d_{d}^2 + d_s^2)
     + \frac{4}{5} d_{fd}^2
     + \frac{17}{15} d_{pd}^2
     + \frac{10}{3} d_{ps}^2
     + \frac{8}{3} d_{pd} d_{ps} \cos (\eta_{pd}-\eta_{ps})
 \Big]. \label{eq:b0}
 \end{equation}

Equations~(\ref{eq:amat1})-(\ref{eq:z}) define $\Delta \tilde \eta$, provided the reduced matrix elements~(\ref{eq:ddn})
are calculated. The intensities of the fundamental and of the second harmonic are factored out in the coefficients $Z'_{\lambda}$, therefore they cancel out in Eqs.~(\ref{eq:cos}), (\ref{eq:sin})
and the phases $\Delta \tilde \eta$ are independent of the intensities of the harmonics.

Note that the angle-resolved average phase difference $\Delta \tilde \eta$ between one- and two-photon
ionization implies not less than two ionization channels, which is reflected
in the non-vanishing sum over channels in Eq.~(\ref{eq:z}). Therefore $\Delta \tilde \eta$ and its energy derivative, or as we called it, generalized delay, is always angle-dependent.

The coefficients (\ref{eq:z}) are directly related to the anisotropy parameters $\beta_{\lambda}$ in the angular distribution of photoelectrons (\ref{eq:apad1}) written in the form
\begin{equation} \label{eq:apad2}
I(\theta; \phi) = \frac{W_0}{4 \pi} \left( 1 + \sum_{\lambda=1}^4 \beta_{\lambda} P_{\lambda}(\cos \theta)
\right) \,,
\end{equation}
where
\begin{equation} \label{eq:bet}
\beta_{\lambda} = \frac{Z_{\lambda}}{Z_0}
\end{equation}
and $W_0 = I_0 Z_0$.
Substituting Eqs.~(\ref{eq:z}) and (\ref{eq:amat1}), (\ref{eq:amat2}) into (\ref{eq:bet}) one can express
the anisotropy parameters in terms of reduced matrix elements (\ref{eq:ddn}): 

\begin{eqnarray} 
\beta_1 & = & \frac{1}{15 Z_0} \Bigg [ \frac{12 \sqrt{3}}{5} d_{d} d_{fd} 
\cos (\eta_{fd}-\eta_d - \phi ) 
- \, \frac{17\sqrt{2}}{5} d_{d} d_{pd} 
\cos (\eta_{pd}-\eta_d - \phi )  \nonumber \\
&  &  - \, 4 \sqrt{2} \, d_{d} d_{ps} 
\cos (\eta_{ps}-\eta_d - \phi )  +  4 \, d_{s} d_{pd} \cos (\eta_{pd}-\eta_s - \phi) \nonumber \\
 &  & + \, 10 \, d_{s} d_{ps} \cos (\eta_{ps}-\eta_s - \phi) \Bigg ] , \label{eq:beta1} \\
 \beta_2 & = & \frac{1}{210 Z_0} \Bigg[  70 d_d^2 + \frac{32}{5} \, d_{fd}^2 + \frac{49}{15} d_{pd}^2 + \frac{140}{3} d_{ps}^2 
  - \, 140 \sqrt{2} \, d_{d}d_{s} 
 \cos (\eta_d-\eta_s)  \nonumber \\
     & &  - \, \frac{48 \sqrt{6}}{5} d_{fd} d_{pd} \cos (\eta_{fd}-\eta_{pd}) - 12 \sqrt{6} \, d_{fd} d_{ps} \cos (\eta_{fd}-\eta_{ps}) \nonumber \\
     & &   + \, \frac{112}{3}  d_{pd} d_{ps} \cos( \eta_{pd}-\eta_{ps}) \Bigg] ,  \label{eq:beta2} \\    
 \beta_3 & = & \frac{1}{30 Z_0} \Bigg[ \frac{16 \sqrt{3}}{5} d_{d} d_{fd} \cos( \eta_{fd}-\eta_d - \phi)
     - \frac{6 \sqrt{2}}{5} d_{d} d_{pd} \cos (\eta_{pd}-\eta_d - \phi) \nonumber \\
     & & - \, 12 \sqrt{2} \, d_{d} d_{ps} \cos (\eta_{ps}-\eta_d - \phi) - 
    4 \sqrt{6} \,d_{s} d_{fd} \cos (\eta_{fd}-\eta_s - \phi) \Bigg] , \label{eq:beta3} \\
 \beta_4 & = &\frac{4}{35 Z_0}            \left[\frac{1}{5} d_{fd}^2- \frac{\sqrt{2}}{5 \sqrt{3}} d_{fd} d_{pd} 
   \cos (\eta_{fd}-\eta_{pd}) - \frac{2 \sqrt{2}}{\sqrt{3}} d_{fd} d_{ps} \cos( \eta_{fd}-\eta_{ps}) \right] \,. \label{eq:beta4}
 \end{eqnarray}

The terms containing cosine functions describe the contribution from the phase difference between two-photon and single-photon ionization channels.
For example, $\cos (\eta_{fd}-\eta_d - \phi )$ in Eq.~(\ref{eq:beta1}) corresponds to the phase difference between $p\to d\to f$ two-photon-ionization (TPI) and $p\to d$ single-photon-ionization (SPI) channels as well as the relative phase of the harmonics $\phi$. Note that $\beta_2$ and $\beta_4$ do not depend on $\phi$.

\begin{figure}[bt]
    \centering
    \includegraphics[width=3.375 in]{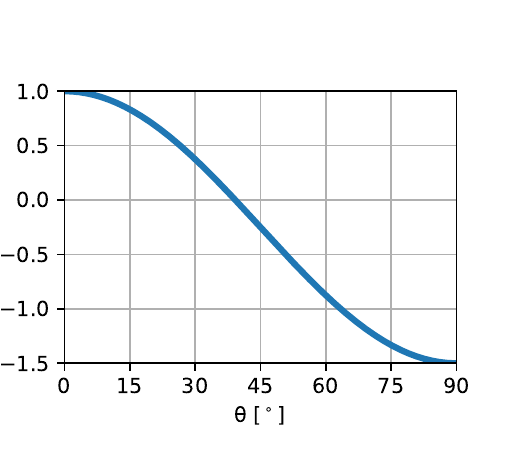}
    \caption{\label{fig:S1} Plot of $(5 \cos 2 \theta -1)/4$ as a function of $\theta$.}
\end{figure}

It immediately follows from Eqs.~(\ref{eq:cos}), (\ref{eq:sin})
\begin{equation} \label{eq:tan}
\tan \Delta \tilde \eta = \frac{{\rm Im} Z'_1 P_1(\cos \theta) 
+ {\rm Im} Z'_3 P_3(\cos \theta)}{ {\rm Re} Z'_1 P_1(\cos \theta) + {\rm Re} Z'_3 P_3(\cos \theta)}
= \frac{ {\rm Im} Z'_1 + f(\theta){\rm Im} Z'_3}{ {\rm Re} Z'_1  + f(\theta) {\rm Re Z'_3}} \,,
\end{equation}
where the function $f(\theta) = \frac{1}{4} \left[5 \cos (2\theta) -1 \right]$ is displayed in Fig.~\ref{fig:S1}.
This qualitatively explains the quasi-cosine shape, and monotonic dependence of $\Delta \tilde \eta$ on $\theta$
(Fig.~\ref{fig:2} C-E).
$Z'_{\lambda}$ are independent of $\phi$, so that $\Delta \tilde \eta$ and the generalized delay are independent of
the relative phase between the harmonics.

The functional form of Eq.~(\ref{eq:tan}) is very general and valid, within the perturbation theory and the dipole approximation, for randomly oriented atoms and molecules, provided corresponding expressions for the coefficients $Z_{\lambda}$ in terms of the ionization amplitudes are used. Moreover, it holds for circularly polarized collinear photon beams (except for chiral targets), provided the angle $\theta$ is measured from the direction of the beam propagation.
 
Expression (\ref{eq:tan}) may be written in an equivalent form in terms of $\theta$-independent ``average partial'' TPI-SPI phase differences $\Delta \tilde \eta_\lambda$ ($\lambda = 1, \, 3$). Indeed, we can write Eq.~(\ref{eq:apad2}) explicitly in the form 
[see Eqs.~(\ref{eq:mod}) and ~(\ref{eq:summation})]
\begin{eqnarray}
  \label{eq:PAD2b}
 	4 \pi W_0^{-1} I(\theta;\phi) & = & [1+\beta_2P_2(\cos\theta)+\beta_4P_4(\cos\theta)]  \nonumber \\
  & & ~~~~~~~	+[\gamma_1\cos(\phi - \Delta \tilde \eta_1)P_1(\cos\theta)+
  \gamma_3\cos(\phi - \Delta \tilde \eta_3)P_3(\cos\theta)] \nonumber \\
 	& = & A_0(\theta) + A(\theta)\cos [\phi - \Delta \tilde \eta \left( \theta \right)]  \,.
\end{eqnarray}
Here the prefactors $\gamma_1$, $\gamma_3$ and phases $\Delta \tilde \eta_1$, $\Delta \tilde \eta_3$ are independent
of $\theta$ and $\phi$. It follows from~(\ref{eq:PAD2b}) that
$\Delta \tilde \eta \left( \theta \right)$ can be viewed as a ``vectorial" average of $\Delta \tilde \eta_1$ and $\Delta \tilde \eta_3$ with weights $\gamma_1P_1(\cos\theta)$ and $\gamma_3P_3(\cos\theta)$, in the sense that $\Delta \tilde \eta \left( \theta \right)$ is the directional angle of vector 
 \begin{equation}
 \label{eq:eq39}
 \left(\gamma_1P_1(\cos\theta) \cos \Delta \tilde \eta_1+\gamma_3P_3(\cos\theta) \cos \Delta \tilde \eta_3, \gamma_1P_1(\cos\theta) \sin \Delta \tilde \eta_1+\gamma_3P_3(\cos\theta) \sin \Delta \tilde \eta_3\right),	
 \end{equation}
satisfying,
\begin{equation}
 \label{eq:eq49}
 	\tan \Delta \tilde \eta \left( \theta \right) = \frac{\gamma_1P_1(\cos\theta) \sin \Delta \tilde \eta_1 + \gamma_3P_3(\cos\theta) \sin \Delta \tilde \eta_3} {\gamma_1P_1(\cos\theta) \cos \Delta \tilde \eta_1 + \gamma_3P_3(\cos\theta) \cos \Delta \tilde \eta_3}.
\end{equation}

There are simple relations between the ``average partial'' TPI-SPI phase differences and parameters of Eq.~(\ref{eq:tan}):
\begin{equation} \label{eq:match1}
 \sin \Delta \tilde \eta_\lambda =\frac{{\rm Im} \, Z'_\lambda}{\left| Z'_\lambda \right|}, \quad 
 \cos \Delta \tilde \eta_\lambda =\frac{{\rm Re} \, Z'_\lambda}{\left| Z'_\lambda \right|}, \quad 
 (\lambda = 1, \, 3)
\end{equation}
and also
\begin{equation} \label{eq:match2}
\gamma_1 = \left| Z'_3 \right|^{-1}, \quad \gamma_3 = \left| Z'_1 \right|^{-1} \,.
\end{equation}
As stated above, we can use the fact that the parity of Legendre Polynomials obeys $P_n(-x) = (-1)^n P_n(x)$, so that the vector defined by Eq.~(\ref{eq:eq39})
changes sign upon performing the substitution $\theta \rightarrow \pi - \theta$, i.e., the two halves of the VMI image oscillate in antiphase: $\Delta \tilde \eta \left( \theta \right) = \Delta \tilde \eta \left( \pi - \theta \right) + \pi$.

%\bibliography{Wigner_PRX.bib}

\end{document}